# Quantum plasmon effects in epsilon-near-zero metamaterials

M. Moaied,[1,2,*] M. M. A. Yajadda,[2] and K. Ostrikov[1,2,3]

[1] *School of Physics, The University of Sydney, Sydney NSW 2006, Australia*

[2] *Commonwealth Scientific and Industrial Research Organisation (CSIRO), PO Box 218, Lindfield NSW 2070, Australia*

[3] *School of Chemistry, Physics and Mechanical Engineering, Queensland University of Technology, Brisbane QLD 4000, Australia*

* m.moaied@physics.usyd.edu.au

# ABSTRACT

Dispersion properties of metals and propagation of quantum bulk plasmon in the high photon energy regime are studied. The nonlocal dielectric permittivity of a metal is determined by the quantum plasma effects and is calculated by applying the Wigner equation in the kinetic theory and taking into account the electron-lattice collisions. The properties of epsilon-near-zero material are investigated in a thin gold film. The spectrum and the damping rate of the quantum bulk plasmon are obtained for a wide range of energies, and the electron's wave function is analytically calculated in both classical and quantum limits. It is shown that the quantum bulk plasmons exist with a propagation length of $1-10$ nm, which strongly depends on the electron energy. The propagation length is found to be much larger than the propagation length in the classical regime which is comparable to the atomic radius and the average inter particle distance. It is found that the spatial localization of the electron wave function is extended due to the quantum effects. Also, it is shown that the damping of electromagnetic waves decreases when the energy of the photon decreases which is opposite to the conclusions obtained from the classical Drude model. The importance of the quantum effects in the development of next-generation metamaterials is also discussed.

# I. INTRODUCTION

The field of quantum plasmonics is growing and motivated by its applications in technology including the investigation of the quantum properties of light and its interaction with matter at the nanoscale.[1-4] The excitation of bulk plasmons as quasi-particles in metals is an interesting phenomenon, particularly, the behaviour of waves in these materials. The bulk plasmons can be used in a wide range of applications, including epsilon-near-zero (ENZ) materials,[5,6] solar cells,[7-10] nanoislanded plasmonic arrays,[11,12] and the development of the ultrafast (e.g., x-ray) laser sources in metallic nanostructures on the typical time scale of the plasmon phenomena.[13,14]

In recent years, ENZ materials have been investigated both theoretically[15-18] and experimentally.[19] In most of the theoretical works, the dielectric permittivity of the metal is described by the classical Drude model. The experimental results[19] have shown that the dielectric permittivity of gold has ENZ property for photon energies beyond $2.4\,\text{eV}$ (corresponding to $517\,\text{nm}$), while ENZ occurs only at the photon energy $2.91\,\text{eV}$ (corresponding to $426\,\text{nm}$) when the classical Drude model[18] is used. Also, the imaginary part of the permittivity, obtained from the experimental data,[19] is higher than the values predicted in the classical Drude model beyond $2\,\text{eV}$ (corresponding to $620\,\text{nm}$), which shows a higher intrinsic loss in gold.

In fact, there is a discrepancy between the classical Drude model and the experimentally measured data for the permittivity of metals at the high photon energies (HPEs). It has shown that the alkali metals can be described by the classical Drude model up to the ultraviolet range.[20,21] In one of the recent works,[22] there is also a difference in spectral shape of a metamaterial (which is designed by composing alternating layers of Ag and SiN) between the measurement and simulation at HPEs beyond $3.1\,\text{eV}$ (corresponding to $400\,\text{nm}$).

Although the plasmon excitation has been extensively studied in the classical case, the quantum case requires further investigations. Considering the density of electronic states in metals at room temperature, the quantum effects cannot be ignored.[23] Understanding the behavior of the plasmonic waves in metals is important, since the quantum effects arise from the quantum nature of the free charge carriers and the dynamic response of these structures to the self-consistent electromagnetic fields. Therefore, it is necessary to use quantum mechanics to model the transport of charge carriers in metallic nanostructures such as metallic nanowires,[24,25] quantum wells,[26] surface plasmon polaritons waveguides,[27,28] metallic nanoparticles,[29,30] etc.

In this work, the quantum bulk plasmons (QBP) are investigated theoretically. The quantum dielectric permittivity (QDP) of metals as a quantum plasma with free mobile electrons and background immobile ions is obtained. It is shown that ENZ property of gold at HPEs beyond $2.91\,\text{eV}$ (corresponding to $426\,\text{nm}$) is determined by the quantum effects. We also show how the loss is decreased in gold when the photon energy decreases. In Sec. II, QDP of metals is obtained by applying the kinetic theory of the quantum plasmas using the Wigner equation while taking into account collisions between free charge carriers and the lattice. In Sec. III, the dispersion relation and damping QBP are investigated, and the decay length of QBP and the electron's wave function are shown. In Sec. IV, the propagation of transverse magnetic (TM) plane wave in a thin film of gold is simulated.

## II. KINETIC THEORY IN QUANTUM PLASMA

Investigating the properties of plasmonic waves in metals as a quantum plasma requires solving the kinetic and Maxwell's equations to describe the excitation of the charged particles by light. We shall limit our analysis to the case of the isotropic collisions in the absence of external electric and magnetic fields. We note that in this model calculation, the interband transitions in metals have not been taken into account.[31] The kinetic equation for the distribution function of free electrons, $f$, in metals with ion background governed by the Wigner equation with the collision term as

$$\frac{\partial f}{\partial t} + \mathbf{v} \cdot \frac{\partial f}{\partial \mathbf{r}} - \frac{iem^3}{(2\pi)^3 \hbar^4} \iint d^3s \, d^3\mathbf{v}' \exp\left[i\frac{m}{\hbar}(\mathbf{v}-\mathbf{v}')\cdot \mathbf{s}\right] \\ \times \left[\phi\left(\mathbf{r}+\frac{\mathbf{s}}{2},t\right) - \phi\left(\mathbf{r}-\frac{\mathbf{s}}{2},t\right)\right] f(\mathbf{v}') = \left(\frac{\partial f}{\partial t}\right)_{col}, \quad (1)$$

where $e$ is the electron charge, $m$ is the electron mass, $\phi$ is the potential at the position of the electron, $\hbar$ is the reduced Planck constant, $(\partial f/\partial t)_{col} = -\nu(f-f_0)$ is the collision integral defined by the Bhatnagar-Gross-Krook (BGK) model,[32] and $\nu$ is the velocity-independent and constant collision frequency describing the electron-lattice collisions, which occurs due to lattice vibrations (phonons) and dominates other collisions (e.g., electron-electron).

In the kinetic theory, the quantum effects become important when the Fermi energy, $E_F = (3\pi^2 n_0)^{2/3} \hbar^2 / 2m$, exceeds the electrons thermal energy, $T_e = \hbar^2/2m\lambda_B^2$,[23,32] where $n_0$ is the number density of electrons and $\lambda_B$ is the thermal de Broglie wavelength of electrons, which quantifies the extension of electron's wave function due to the quantum uncertainty. This is the condition of fully degenerate electron plasmas which is satisfied up to the temperature of $10^4 \mathrm{K}$ for metals.[32] In other words, quantum effects play an important role when $\lambda_B$ is larger than the average inter-particle distance ($n_0^{-1/3}$). The solution of Eq. (1) in the stationary state is the equilibrium Fermi distribution function

$$f_0 = \frac{2}{(2\pi\hbar)^3}\left\{1+\exp\left[\frac{(p^2/2m)-E_F(n_0)}{T_e}\right]\right\}^{-1}, \quad (2)$$

where $p$ is the momentum of electrons.

We introduce a perturbed field into the equilibrium of the metal as $\delta f = f - f_0$ where $\delta f \ll f_0$. Accordingly, $\delta f$ and $\phi$ are proportional to $\exp(-i\omega t + i\mathbf{k}\cdot\mathbf{r})$, where $\omega$ is the frequency and $\mathbf{k}$ is the wave vector of the perturbation, and the associated fields are small, from Eq. (1) we obtain the first-order linear kinetic equation

$$\begin{aligned}
\left[(\omega+i\nu)-\mathbf{k}\cdot\mathbf{v}\right]\delta f &+ i\nu\frac{2E_F}{3n_0}\int \delta f\, d\mathbf{p} = \\
&-\frac{em^3}{(2\pi)^3\hbar^4}\iint d^3s\, d^3v' \exp\left[i\frac{m}{\hbar}(\mathbf{v}-\mathbf{v}')\cdot\mathbf{s}\right] \\
&\times\left[e^{i\mathbf{k}\cdot\mathbf{s}/2}-e^{-i\mathbf{k}\cdot\mathbf{s}/2}\right]f_0(\mathbf{v}')\phi(\omega,\mathbf{k}),
\end{aligned} \quad (3)$$

that by integrating over $\mathbf{s}$ and $\mathbf{v}'$ spaces, gives

$$\delta f = -\frac{e}{\hbar}\frac{f_0\left(\mathbf{v}+\frac{\hbar\mathbf{k}}{2m}\right)-f_0\left(\mathbf{v}-\frac{\hbar\mathbf{k}}{2m}\right)}{\left[(\omega+i\nu)-\mathbf{k}\cdot\mathbf{v}\right]}\phi(\omega,\mathbf{k}) - i\nu\frac{2E_F}{3n_0}\frac{\int \delta f\, d\mathbf{p}}{\left[(\omega+i\nu)-\mathbf{k}\cdot\mathbf{v}\right]}, \quad (4)$$

and substituting Eq. (4) into the charge and the current densities (which defined as $\rho = e\int \delta f\, d\mathbf{p}$ and $\mathbf{j} = (e/m)\int \mathbf{p}\, \delta f\, d\mathbf{p}$, respectively) in Maxwell's equations and integrating over $\mathbf{v}$ space, we finally obtain the susceptibility of electrons in metals as

$$\chi(\omega,k) = \frac{\dfrac{3\omega_p^2}{4k^2 V_F^2} \left\{ \begin{aligned} & 2 - \frac{m}{\hbar k^3 V_F}\left[k^2 V_F^2 - \left(\omega+i\nu+\frac{\hbar k^2}{2m}\right)^2\right]\ln\left|\frac{\omega+i\nu-kV_F+\frac{\hbar k^2}{2m}}{\omega+i\nu+kV_F+\frac{\hbar k^2}{2m}}\right| \\ & + \frac{m}{\hbar k^3 V_F}\left[k^2 V_F^2 - \left(\omega+i\nu-\frac{\hbar k^2}{2m}\right)^2\right]\ln\left|\frac{\omega+i\nu-kV_F-\frac{\hbar k^2}{2m}}{\omega+i\nu+kV_F-\frac{\hbar k^2}{2m}}\right| \end{aligned} \right\}}{\left[1 - \dfrac{i\nu}{2kV_F}\ln\left|\dfrac{\omega+i\nu+kV_F}{\omega+i\nu-kV_F}\right|\right]}, \qquad (5)$$

where $\omega_p = \sqrt{4\pi n_0 e^2/m}$ and $V_F = \sqrt{2E_F/m}$ are the plasma frequency and the Fermi velocity of the electrons, respectively, and $k = |\mathbf{k}|$ is the wave number. In the limit of $k \to 0$, Eq. (5) reduces to the classical model $\chi(\omega) = -\omega_p^2/[\omega(\omega+i\nu)]$ in metals. In other words, the classical limit gives the condition $kV_F, \hbar k^2/2m \ll \omega$. Therefore, we have the limits $k \ll \omega/V_F$ and $k \ll \sqrt{2m\omega/\hbar}$ in the classical regime. One can thus expect stronger quantum effects at higher $k$, which leads to spatial nonlocality of the metal.[29,33,34]

We consider the permittivity of quantum plasma as $\varepsilon = \varepsilon_\infty + \chi(\omega,k)$ where $\varepsilon_\infty$ is the offset value of the permittivity and $\chi(\omega,k)$ is given by Eq. (5). Figure 1 shows the $k$-dependent QDP for gold with $\varepsilon_\infty = 9.6$, $\hbar\omega_p = 9.03$ eV, $V_F = 1.4\times 10^6$ m/s, and $\hbar\nu = 70.3$ meV [18,23] at various frequencies. Figure 1(a) shows that the QDP is negative in the space between the dashed lines obtained from $\mathrm{Re}[\varepsilon(\omega,k)] = 0$. In fact, the dashed lines explain ENZ properties of gold when QBP resonances with specified frequencies and wavenumbers are expected. One can see two different frequencies for the given $k$ under the ENZ conditions. The low-frequency mode starts from the infrared and the terahertz ranges and increases monotonically with $k$. While, the high-frequency mode starts from the value of 2.91 eV (corresponding to 426 nm) in the visible range and increases with $k$. These two frequencies are merged together at the energy of 4.7 eV (corresponding to 264 nm) and the wavenumber of 4.6 nm$^{-1}$ in the ultraviolet range. Figure 1(b) shows that the low-frequency mode of QBP cannot propagate, because of strong absorption in metals, compared to the high-frequency mode with less absorption.

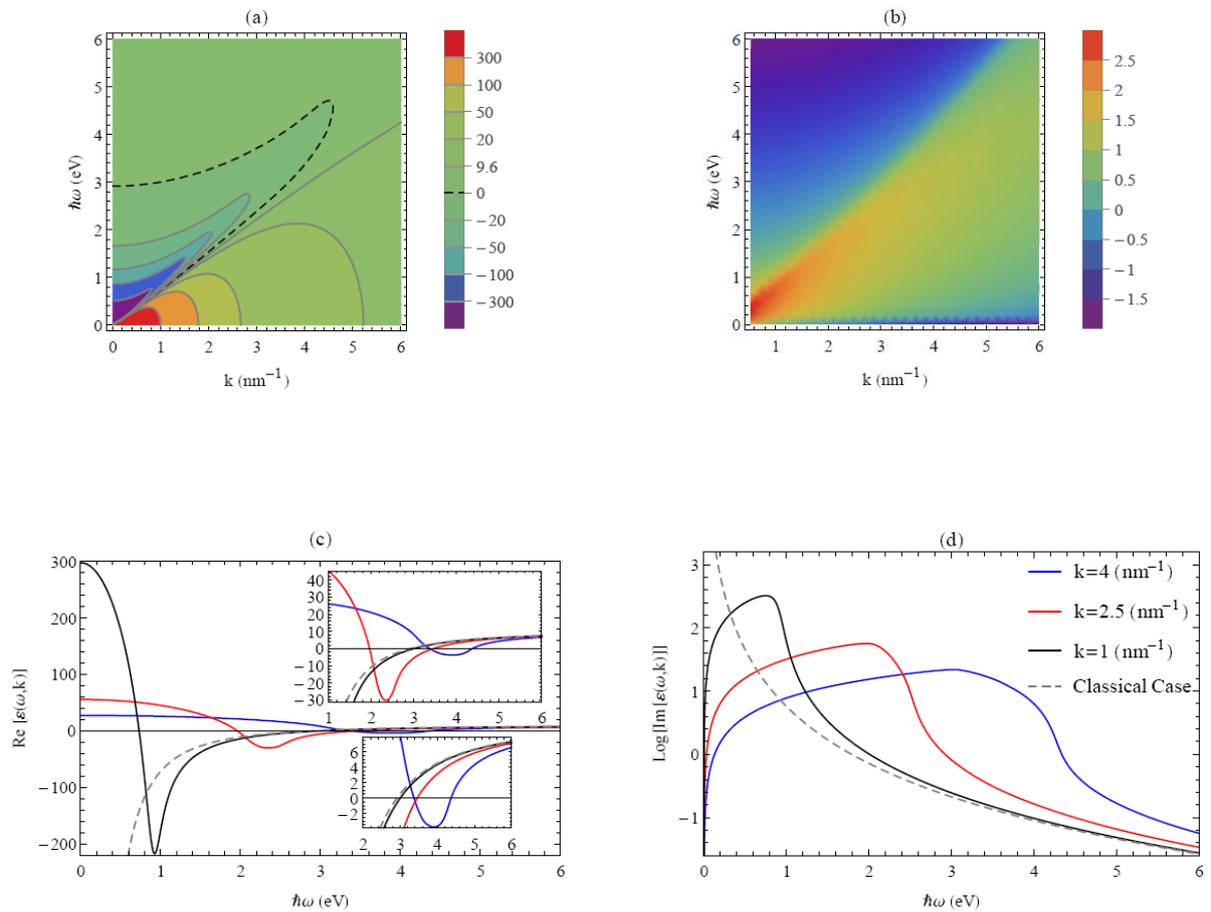

FIG. 1. The frequency- and $k$-dependent QDP of gold; (a) Re[$\varepsilon(\omega,k)$] (dashed line, Re[$\varepsilon(\omega,k)$] = 0); (b) log[Im[$\varepsilon(\omega,k)$]]. (c) and (d) are real and imaginary parts of QDP, respectively, for various $k$.

Figures 1(c) and 1(d) show the real and the imaginary parts of QDP, respectively, for the wavenumber values of 1, 2.5, and 4 nm$^{-1}$. The energy of the excitation $\left(\text{Re}[\varepsilon(\omega,k)]=0\right)$ at the high-frequency mode increases when the wavenumber of QBP increases while in the classical limit $k \ll 1$ nm$^{-1}$ there is no change in the excitation energy. One can also see that the absorption in the high-frequency mode increases as the QBP wavenumber increases. In the next section we show that the quantum effects become more important in the excitation of QBP when the energy of the excitation increases.

## III. DISPERSION OF QUANTUM PLASMONS

The dispersion equation for a bulk medium,[32] $\varepsilon(\omega,k)=0$, gives the existence of QBP in metals. Out of all complex-valued solutions of the dispersion equation as $\omega(k) = \omega_0(k) + i\gamma_0(k)$, only those corresponding to the weakly damped waves (i.e., high-frequency mode), with $|\gamma_0| \ll |\omega_0|$, are of physical interest where $\omega_0$ and $\gamma_0$ are the spectra and the damping of QBP which are obtained from the real and the imaginary parts of the dispersion equation, respectively. Therefore, by using an analytical calculation, one obtains the dispersion and the damping of QBP (Fig. 2). This shows that the weakly damped excitations start from the energies around $\hbar\omega_0 \approx \hbar\omega_p/\sqrt{\varepsilon_\infty} = 2.91$ eV with wavenumbers $k \ll 1$ nm$^{-1}$ (Fig. 2(a)) in the classical regime and increase with $k$ (nonlocality) in the quantum regime up to the energy of $4.71$ eV (corresponding to $263$ nm) with the wavenumber $k = 4.35$ nm$^{-1}$.

Figure 2(b) shows that the damping rate is $|\hbar\gamma_0| \approx \hbar\nu/2 = 35.15$ meV in the classical case, taking into account the effect of the electron-lattice collisions. So, there is only the

collisional damping in the classical limit. The damping rate increases up to $69.61\,\text{meV}$ (which is still weak) in the quantum regime. This is due to the Cherenkov absorption[32,35] which is a result of the resonant absorption of QBP energy by electrons and it grows when the quantum effects of the electron's motion become important.

This absorption is caused by the exchange of energy between electrons and QBP. The electric field of QBP accelerates electrons which have a velocity smaller than the phase velocity of QBP. On the other hand, the electron with a velocity higher than the phase velocity of QBP will be slowed down by the electric field of the wave. Therefore, without considering the resonant excitation of electrons, the loss will be less in HPE regime (see Fig. 1(d)). In the quantum limit, the total damping of QBP in metals is a superposition of the Cherenkov and the collisional damping. In HPE regime, one can see that the Cherenkov damping is more effective than the collisional damping.

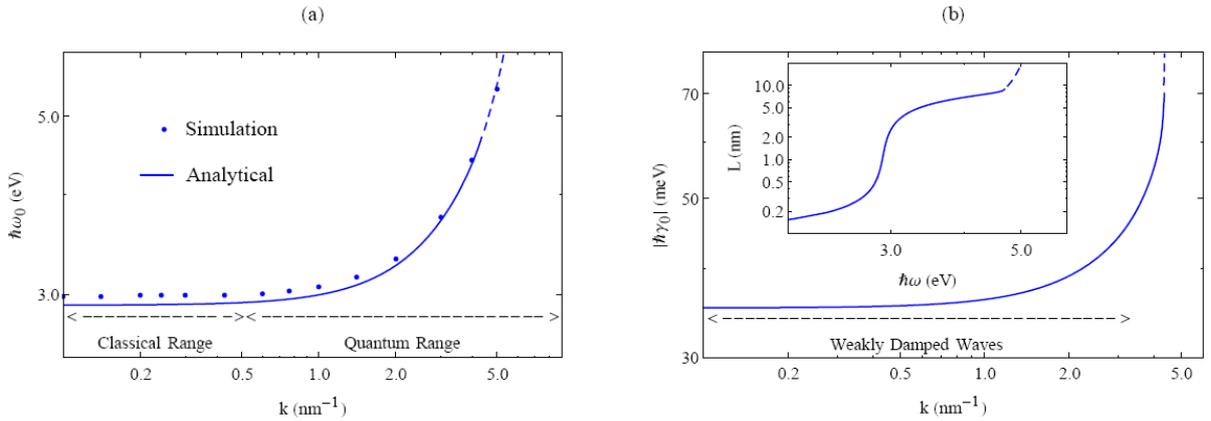

FIG. 2. Dispersion of QBP in gold. (a) spectra (b) damping rate, inset shows the propagation length. Dashed lines correspond to strongly damped waves and dots in Fig. 2(a) correspond to the geometry considered in Sec. IV.

One can also solve the dispersion equation with respect to $k(\omega)=k'(\omega)+ik''(\omega)$ where $k'$ and $k''$ are the real (QBP wavenumber) and the imaginary (decay length) parts of $k$, respectively. The intensity of QBP, $I(r)$, decays in space with the square of the electric field as

$$I(r) = I_0 \left|\exp[ikr]\right|^2 = I_0 \exp[-2k''r], \quad (6)$$

where $I_0$ is the intensity at $r=0$. Therefore, the propagation length of QBP is defined as $L = 1/2k''$ shown in the inset of Fig. 2(b). In the classical energy range (less than 2.91 eV) $L$ is of the order of atomic radius (1.7 Å for gold)[23] and the average inter-particle distance (2.6 Å for gold).[23] The quantum effects become significant when $L$ increases as the energy increases. Therefore, electrons can be found at higher energy levels. From inset of Fig. 2(b), one can see that $L$ increases to the range of 1 to 10 nm in the quantum regime for the weakly damped QBP.

Figure 3 shows the intensity of QBP $I(r)$ and the electron's wave function

$$\Psi(r) = A\exp[ikr] \quad (7)$$

at various energies. Here, $A$ is the amplitude of the electron's wave function. One can see the extension of the electron's wave function in space due to the quantum effects when the energy of QBP increases. For the classical regime, $\hbar\omega \leq 2.91$ eV, this extension is smaller than or of the order of the average inter-particle distance that electrons can be considered as a point and the classical statistical pressure dominates the wave dynamics. Whereas in the opposite case, the quantum effects are important when the wave function extends beyond the average inter-particle distance and the Cherenkov absorption increases in the system.

Therefore, the effect of the quantum tunnelling becomes important and the damping increases in metal.

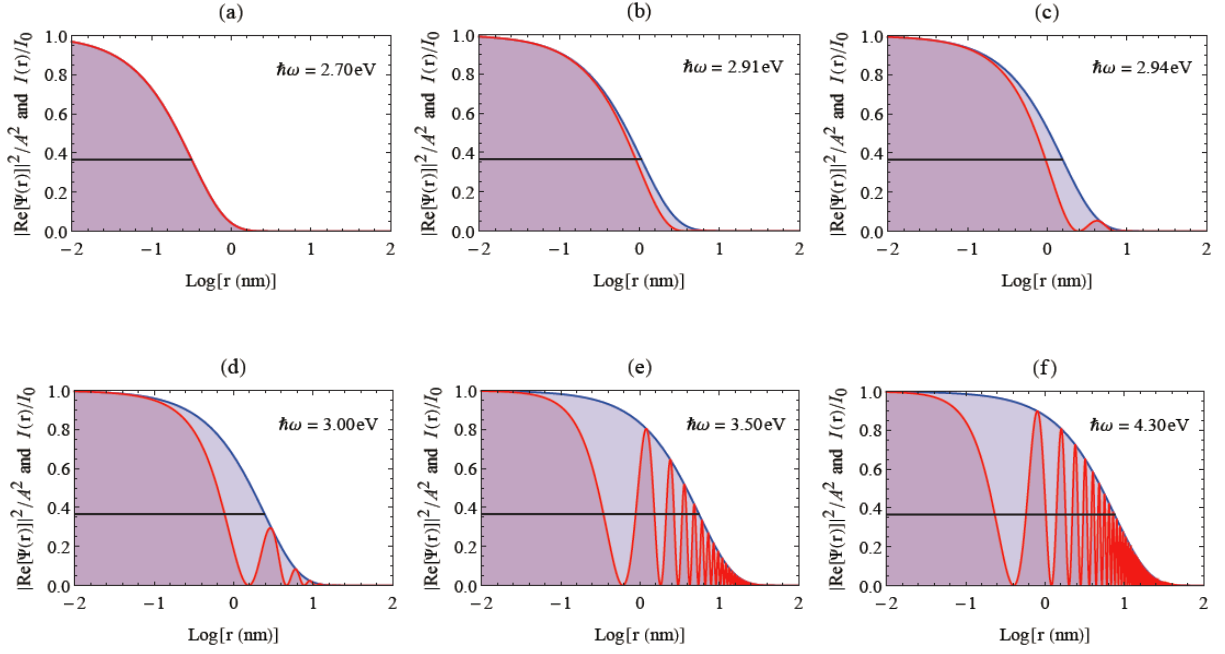

FIG. 3. The intensity of QBP $I(r)$ (blue line) and the real part of the electron's wave function $|\text{Re}[\Psi(r)]|^2$ (red line), obtained from Eqs. (6) and (7), respectively, at various energies. Black horizontal line, the propagation length $L$ (corresponding to the inset of Fig. 2(b)).

## IV. PROPAGATION OF ELECTROMAGNETIC WAVES IN METALS

In this section, we perform numerical calculations of the propagation of electromagnetic waves in the plasmonic structure using COMSOL Multiphysics. Consider the geometry[16,22] shown in Fig. 4(a), which shows an infinitely extended planar slab of gold, with the permittivity defined by the quantum model of the charge carriers as $\varepsilon(\omega, k)$ and the thickness $200\,\text{nm}$ which is deposited on a dielectric with a refractive index of $n_d = 1.5$ and excited by a TM plane wave with the wave vector $\mathbf{k_0}$ and the wavelength $\lambda_0$ in air ($n_a = 1$).

The blue points in Fig. 2(a) show the dispersion of QBP for this geometry, which is close to the analytical results. This shows that the excitations start from the energy around $2.98\,\text{eV}$ (corresponding to $416\,\text{nm}$) and this energy increases as well as analytical results in the quantum regime.

Figure 4(b) shows the absorption of TM waves with the photon energies $2.98\,\text{eV}$ (corresponding to $416\,\text{nm}$), $3.61\,\text{eV}$ (corresponding to $343\,\text{nm}$), and $4.66\,\text{eV}$ (corresponding to $266\,\text{nm}$) in gold. The peak of the absorption corresponds to the excitation of QBP. For the wave with the energy of $2.98\,\text{eV}$, bulk plasmon is excited with $k = 0.27\,\text{nm}^{-1}$ which is in the classical range. While, the photons with higher energies excite QBPs with the wavenumbers $k = 2.78\,\text{nm}^{-1}$ and $k = 4.27\,\text{nm}^{-1}$ at the energies of $3.61\,\text{eV}$ and $4.66\,\text{eV}$, respectively. One can also see that the intensity of the absorption peak increases at higher energies due to quantum effects (e.g., tunnelling and Cherenkov damping in metals).

The 3D distributions of the z component of the magnetic field, $H_z$, shown in Figs. 4(c), (d), and (e), describe the propagation of TM waves at the peak of absorption in Fig. 4(b). The intensity of TM waves is absorbed up to $61\%$, $82\%$, and $94\%$ of the initial value while crossing the plasmonic structure at the photon energies of $2.98\,\text{eV}$, $3.61\,\text{eV}$, and $4.66\,\text{eV}$, respectively. It is important to note that the light absorption in noble metals can occur due to interband transitions as well, which have not been considered in this model calculation.[31]

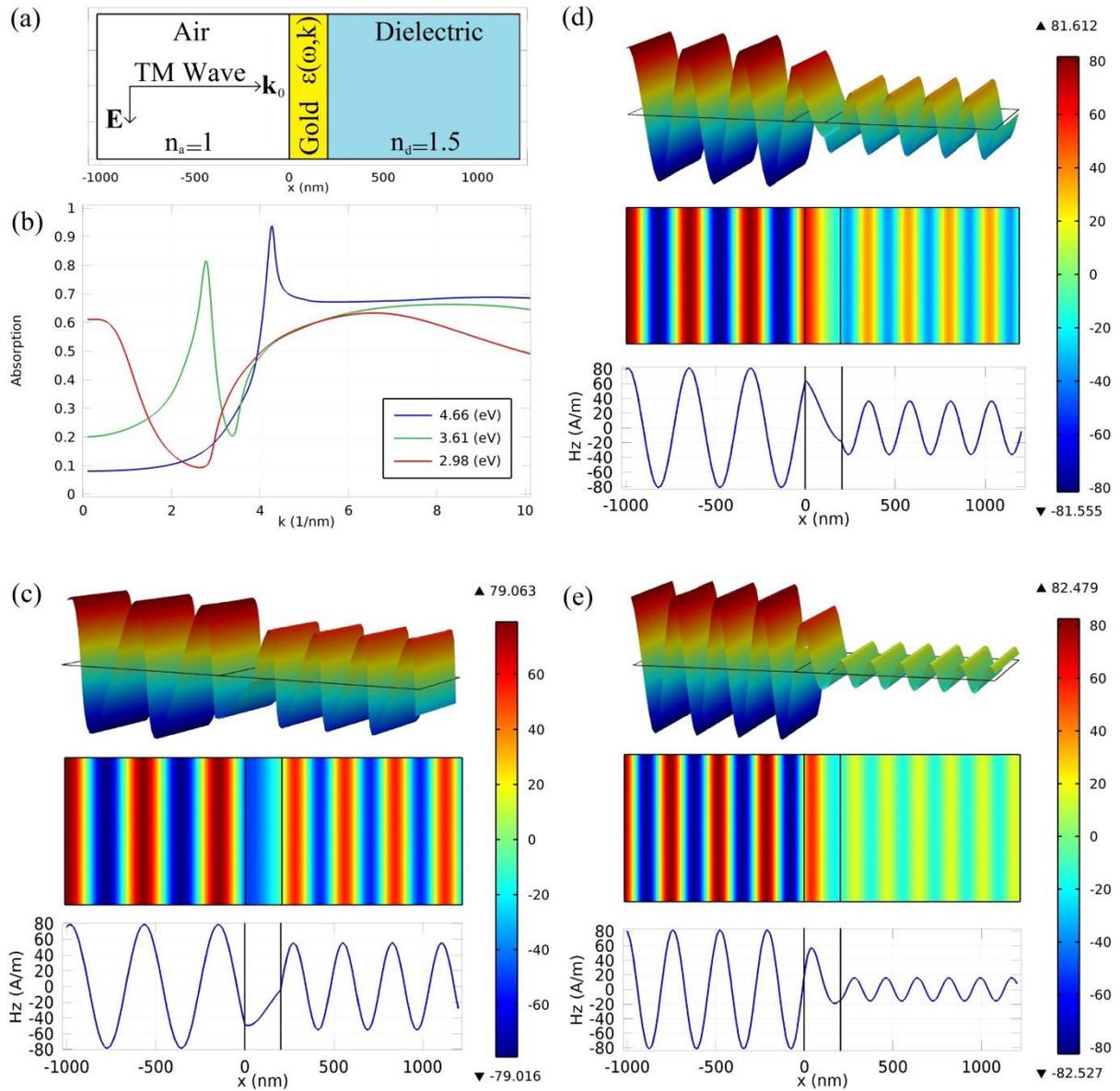

FIG. 4. Transmission of TM plane wave through a planar slab of gold in the quantum regime. (a) geometry (b) absorption spectra of TM waves by quantum plasmas. The z component of the magnetic field, $H_z$, for TM waves at the peak of absorption: (c) $2.98\,\text{eV}$, (d) $3.61\,\text{eV}$, and (e) $4.66\,\text{eV}$.

These results can be used to design customized metamaterials, for example, composed of alternating layers of a metal and a dielectric, where the layers are of a subwavelength thickness.[22] The effective permittivity of this metamaterial equals the geometrically averaged permittivity,[36] $\varepsilon_{av} = \alpha\varepsilon_m + (1-\alpha)\varepsilon_d$, where $\alpha$ is the metal filling fraction and $\varepsilon_m$ and $\varepsilon_d$ are the permittivity of the metal and the dielectric, respectively. Therefore, in the classical limit, the ENZ condition (which shows the fundamental waveguide mode in this structure) for Au-SiN layers with $\varepsilon_d = 5$ for the dielectric permittivity of SiN and the filling fractions $\alpha = 20\%, 30\%,$ and $50\%$ occurs at the photon energies of $1.66\,\text{eV}$ (corresponding to $747\,\text{nm}$), $1.96\,\text{eV}$ (corresponding to $632\,\text{nm}$), and $2.36\,\text{eV}$ (corresponding to $525\,\text{nm}$), respectively. Therefore, the quantum effects become important at the energies higher than the above-mentioned energies. Specially, for the first higher-order waveguide mode, the experimental data[22] shows that the imaginary part of the dielectric permittivity for this mode becomes comparable to that of the fundamental mode, where cannot be accounted by using a classical Drude model. The application of QBP in the metamaterials composed of alternating layers of metal and dielectric is outside the scope of this paper and requires further investigations.

## V. CONCLUSION

In this paper, the quantum effects of bulk plasmon have been studied. The Wigner equation was applied to the kinetic theory of free mobile electrons in metals with background immobile ions where the dominant electron scattering mechanism was the electron-lattice interaction. The nonlocal QDP ($k$-dependence of metal responses) was introduced and the dispersion relation corresponding to the weakly damped QBP in gold was investigated.

The calculations in the quantum regime suggest the possibility of excretion of QBPs in gold with the wavenumber of $k \approx 1\,\text{nm}^{-1}$, while the classical limit shows bulk plasmons with the wavenumber of $k \ll 1 \text{ nm}^{-1}$. These plasmons have the energy from $\hbar\omega_p/\sqrt{\varepsilon_\infty} = 2.91 \text{ eV}$ in the classical regime, which increases to $4.71 \text{ eV}$ in the quantum range. It is found that the spatial localization of the electron wave function is of the order of atomic radius and the average inter particle distance in the classical regime, while it is dramatically increased to $1-10 \text{ nm}$ in the quantum regime due to increase of QBP energy. This can be attributed to the stronger wave-like properties of the electrons in HPE regime.

It was shown that the probability of finding electrons at higher energy levels increases in the excitation of QBP, since their wave functions overlap, and therefore the quantum tunnelling effect increases in metals. Consequently, the losses in metals increase due to the plasma wave dephasing at HPEs.

These results are applicable in thin metal film structures, custom-design metamaterials, solar cells, development of the ultrafast laser sources in metallic nanostructures, and other plasmonic devices operated at high photon energies.

## ACKNOWLEDGMENTS

M. Moaied would like to thank the University of Sydney for receiving an Australian Postgraduate Award and the CSIRO for the financial support through the OCE PhD Scholarship Top-up Scheme. This work was partially supported by the Australian Research Council and CSIRO's Science Leadership Scheme.